\documentclass[aps,prl,twocolumn,superscriptaddress,longbibliography]{revtex4-2}
\usepackage{graphicx}
\usepackage{dcolumn}
\usepackage{bm}
\usepackage{amsmath}
\usepackage{mathrsfs}
\usepackage{hyperref}
\hypersetup{
    colorlinks=true,
    linkcolor=blue,       
    filecolor=magenta,  
    urlcolor=cyan,       
    citecolor=purple,         
    pdftitle={Smart Lagrangian forcing},
    pdfpagemode=FullScreen,
}

\usepackage{xcolor}
\usepackage{float}
\floatstyle{plaintop}
\restylefloat{table}
\makeatletter
\let\newfloat\newfloat@ltx
\makeatother
\usepackage{algorithm}
\usepackage{algpseudocode}
\usepackage{setspace}
\usepackage{lipsum}
\usepackage{subcaption}
\usepackage{booktabs} 
\usepackage{ragged2e}

\def\equationautorefname#1#2\null{%
  Eq.#1(#2\null)%
}

\newcommand\void[1]{}

\begin{document}


\title{Intermittency suppression in turbulence via forced light particles}

\author{André Freitas}
\thanks{These authors contributed equally}
\affiliation{Department of Physics and INFN, University of Rome “Tor Vergata”, Italy}
\email{andre.freitas@roma2.infn.it}
\affiliation{LTCI, Télécom Paris, IP Paris, France}
\author{Xander M. de Wit}
\thanks{These authors contributed equally}
\affiliation{Department of Applied Physics and Science Education, Eindhoven University of Technology, Eindhoven, Netherlands}
\author{Ziqi Wang}
\affiliation{Department of Applied Physics and Science Education, Eindhoven University of Technology, Eindhoven, Netherlands}
\author{Luca Biferale}
\affiliation{Department of Physics and INFN, University of Rome “Tor Vergata”, Italy}
\author{Federico Toschi}
\affiliation{Department of Applied Physics and Science Education, Eindhoven University of Technology, Eindhoven, Netherlands}
\affiliation{Istituto per le Applicazioni del Calcolo “M. Picone”, Consiglio Nazionale delle Ricerche, Rome, Italy}

\date{\today}

\begin{abstract}
We investigate how turbulence is reshaped by the presence of externally forced light particles, using high-resolution direct numerical simulations with four-way coupling. The particles are subject to an oscillatory force that in turn locally affects the fluid flow through momentum exchange at the position of the particles. Since the light particles preferentially concentrate in high vorticity regions, this leads to an intricate preferential turbulence modulation. We show that through this modulation, the forced light particles strongly reduce the intermittency of the flow, shedding new light on the delicate relationship between vortex filaments and turbulence intermittency.
\end{abstract}

\maketitle

\emph{Introduction} – 
Turbulence is a paradigmatic example of a nonlinear, multiscale phenomenon, where energy injected at large scales cascades down to smaller scales and dissipates through viscous effects~\cite{frisch}. The statistical characterization of this cascade—especially its intermittent nature at small scales—remains a central challenge~\cite{frisch1980fully}. Despite decades of research, the origin and control of intermittency and anomalous scaling in fully developed turbulence remain only partially understood.
The problem becomes even richer when one introduces particles into the flow~\cite{coletti_rev}. Finite-size, inertial particles—either heavier or lighter than the fluid—do not simply trace the flow. Instead, they interact with specific structures in turbulence, such as high-strain regions~\cite{bec_effects_2006} or vortex filaments, leading to preferential concentration, clustering, and non-trivial transport properties~\cite{toschi_lagrangian_2009}. Light particles, in particular, are known to accumulate within vortex filaments, a behavior that can be conceptualized as a dynamic trapping mechanism induced by local pressure minima and high enstrophy regions~\cite{toschi_lagrangian_2009_not_review}. Studying turbulence with particles is not only key to understanding these Lagrangian dynamics, but also essential for modeling and controlling real-world systems where such interactions are fundamental—from clouds~\cite{clouds} and sprays~\cite{crowe1998multiphase} to protoplanetary disks~\cite{planetary} and pollutant transport.
At finite concentrations, particles are numerically modeled using different coupling paradigms beyond passive advection (one-way coupling). Two-way coupling accounts for momentum exchange with the fluid~\cite{elghobashi_two-way_1993}, while four-way coupling additionally includes inter-particle collisions to enforce excluded volume and avoid unphysical clustering~\cite{bec_sticky_2013}. This opens a path to the active control or modulation of turbulence via Lagrangian agents—particles that are not only immersed in turbulence but dynamically coupled to it. Previous efforts at turbulence modulation have hitherto mostly focused on Eulerian approaches~\cite{buzzicotti_statistical_2020, si2024manipulatingdirectionturbulentenergy}, with some exploratory studies on Lagrangian approaches applied to simpler flow configurations~\cite{colabrese_smart_2018, Agasthya2022} and heavy particle settling~\cite{Tom_Carbone_Bragg_2022,Bhattacharjee_Tom_Carbone_Bragg_2024}. 

\begin{figure*}[htb]
    \centering
    \includegraphics[width=0.95\linewidth]{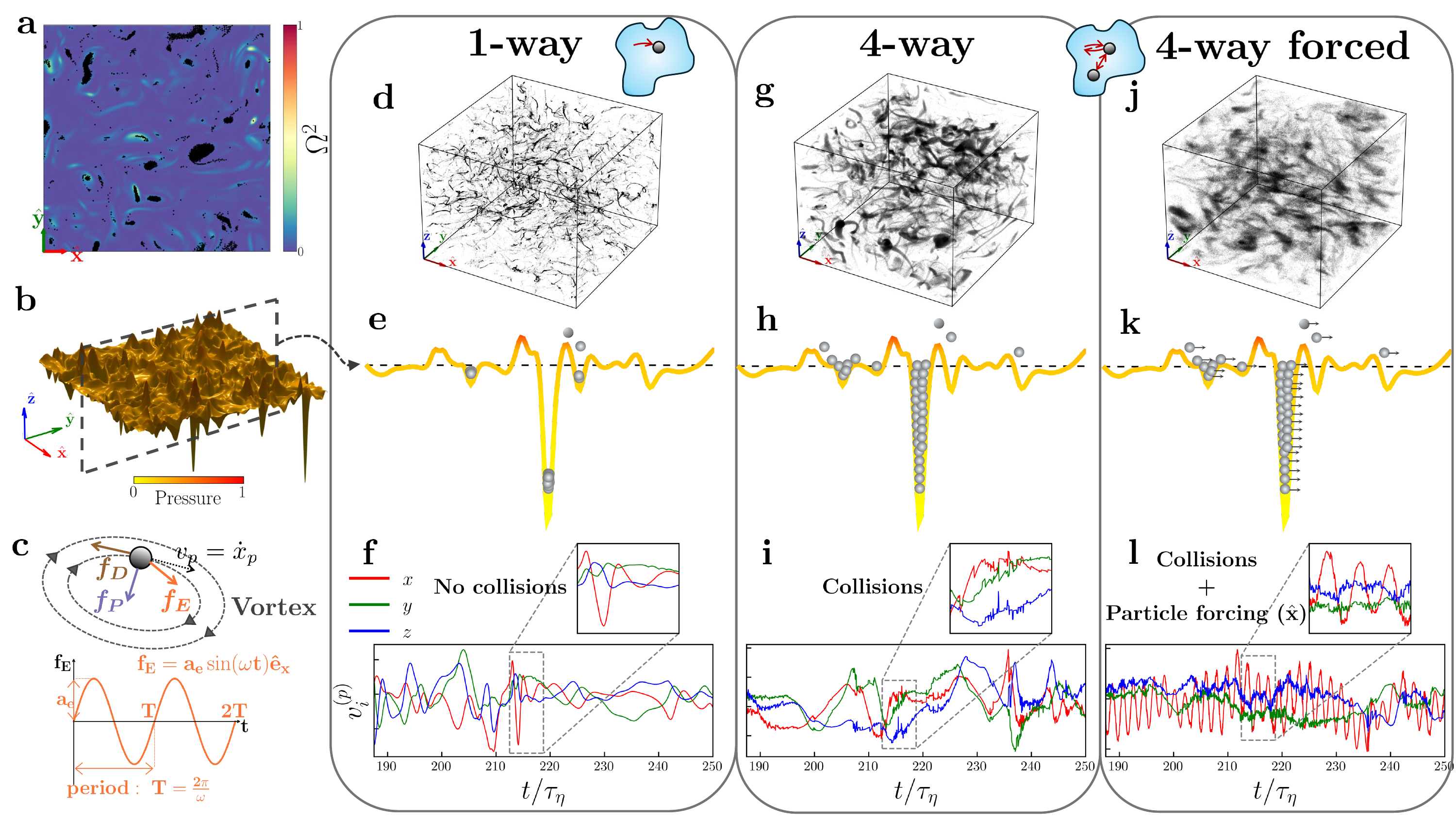}
    \caption{\justifying \textbf{Problem setup}. (a) 2D slice of the enstrophy field $\Omega^2$ with particles in black, showing preferential concentration in high-enstrophy regions. (b) Pressure field at fixed $z$. (c) Forces on a trapped particle: Stokes drag $f_D$, added mass and pressure gradient $f_P$, and external forcing $f_E = a_e \sin(\omega t) \hat{\mathbf{e}}_x$. We compare three cases: 1-way coupling (particles advected by the flow, no feedback), 4-way coupling without forcing (includes momentum exchange and collisions), and 4-way coupling with external particle forcing. (d,g,j): 3D particle distributions. Clustering is strongest in the 1-way case, where particles collapse into vortex filaments; in the 4-way cases, collisions limit accumulation, and forcing enables escape, leading to more diffuse distributions. (e,h,k): slices of the pressure field, illustrating how particle positions relate to pressure minima in each regime. (f,i,l): particle velocity components, highlighting the effects of collisions and forcing.}
    \label{fig:intro_panel}
\end{figure*}

In this work, we explore how homogeneous isotropic turbulence (HIT) is modified when light particles are actively driven by an external oscillatory force. We employ high-resolution direct numerical simulations with four-way coupling to capture the full interplay between particles, fluid, and feedback. From an experimental perspective, such oscillatory forcing can be realized through several protocols such as magnetic fields driving magnetized particles~\cite{wu2025trackingrotationlightmagnetic,wang_stochastic_2025}, acoustic waves~\cite{crum1970motion} or mechanical shaking~\cite{seropian2023effect}.
Our central finding is that smart Lagrangian forcing can suppress the intermittency of the flow, leading to a measurable modulation of turbulence. This modulation arises from a combination of resonant effects that depend on the amplitude and frequency of the external forcing, feedback from particle-fluid interactions, and a dynamic trapping-and-release mechanism induced by the oscillatory nature of the forcing (previously studied by Wang et al.~\cite{wang_localization-delocalization_2024}). Additionally, volume exclusion due to particle collisions plays a key role: as particles fill vortex filaments, they increase the effective volume fraction influenced by the forcing, thereby amplifying its impact on the turbulent structures. The problem setup is illustrated in \autoref{fig:intro_panel}.
\raggedbottom

\emph{Fluid, particle phases and coupling} – 
The turbulent carrier flow is governed by the incompressible Navier–Stokes equations:
\begin{equation}
    \frac{D\mathbf{u}}{Dt} = -\nabla p + \nu \nabla^2 \mathbf{u} + \mathbf{f} + \mathbf{f}_p\,, 
    \hspace{0.7cm} \nabla \cdot \mathbf{u} = 0\,,
\end{equation}
where $\mathbf{u}(\mathbf{x}, t)$ is the fluid velocity, $p$ is the pressure, $\nu$ is the kinematic viscosity, and $\mathbf{f}$ represents large-scale forcing $\hat{\mathbf{f}}(\mathbf{k})=\epsilon \hat{\mathbf{u}}(\mathbf{k}) / \sum_{k_f \leq|\mathbf{k}|<k_f+1}|\hat{\mathbf{u}}(\mathbf{k})|^2$ acting on low wavenumbers, that maintains a constant energy injection rate $\epsilon$. The particle feedback force $\mathbf{f}_p$ accounts for the momentum exchange between the particles and the fluid. Simulations are performed via a pseudospectral solver with periodic boundary conditions, ensuring full resolution of the smallest scales, represented by the Kolmogorov length $\ell_\eta$, with $k_{\max} \ell_\eta \gtrsim 3$. In particular, we use a resolution of $N^3 = 256^3$ and $Re_{\lambda} = 87$. Complementary results for decaying turbulence are reported in the End Matter. Particles are modeled as spherical point particles using the Maxey–Riley equations~\cite{maxey_equation_1983} in the small-size, high-density-contrast limit. We retain only the Stokes drag and added-mass terms, yielding the following evolution equations for particle position $\mathbf{X}(t)$ and velocity $\mathbf{V}(t)$:
\begin{subequations}
\begin{align}
    \frac{d\mathbf{X}}{dt} &= \mathbf{V} + \text{(collisions)}, \\
    \frac{d\mathbf{V}}{dt} &= \beta \frac{D\mathbf{u}}{Dt} - \frac{1}{\tau_p}(\mathbf{V} - \mathbf{u}) + \bm{f}_E,
\end{align}
\end{subequations}
where $D\mathbf{u}/Dt$ is evaluated at the particle location, $\tau_p$ is the particle response time, $\beta = 3/(1 + 2\rho_p/\rho_f)$ captures the density ratio between particle and fluid, and $\bm{f}_E = a_e\sin(\omega t) \hat{\mathbf{e}}_x$ is the external oscillatory forcing applied to the particle. We focus on the limit $\beta = 3$ relevant for bubbles, and fix $St = \tau_p / \tau_\eta = 1$ throughout (unless mentioned otherwise). Interpolation of the fluid velocity and its material derivative onto the particle position is done using a B-spline interpolation~\cite{michel_interp}. Four-way coupling is implemented to capture the full dynamics of the fluid–particle system as described in detail in Ref.~\cite{de_wit_efficient_2024}. This includes (i) the advection of particles by the flow (one-way), (ii) the feedback of particles onto the fluid through local momentum exchange (two-way), and (iii) hard-sphere particle–particle interactions enforcing excluded volume constraints (four-way). The fluid velocity at the particle position is taken as the local (actual) velocity field, an approximation validated in our previous work~\cite{de_wit_efficient_2024}, where the self-induction effect was shown to be negligible under the conditions considered in this work. The feedback force from the particle phase is included in the fluid equations as:
\begin{equation}
    \mathbf{f}_p(\mathbf{x},t) = \sum_i \left[ \frac{D\mathbf{u}}{Dt} - \frac{\rho_p}{\rho_f} \frac{d\mathbf{V}_i}{dt} + \frac{3}{2\beta}f_{E} \right] \mathscr{V}_p \delta(\mathbf{x} - \mathbf{X}_i(t)),
\end{equation}
where $\mathscr{V}_p$ is the particle volume. A detailed derivation of this feedback term is given in the End Matter. The delta function is discretized by linear volume weighting to the eight surrounding grid nodes and subsequently projected to divergence-free form. We have verified that for this case using different interpolation kernels does not influence the reported results.
Collisions are handled via an efficient algorithm based on spatial boxing and resolution of strongest overlaps per time step, ensuring stability and low-overlap. This setup enables high-resolution, large number of particle simulations in regimes of strong particle clustering and significant feedback onto the turbulence.

\begin{figure}[tb]
    \centering
    \includegraphics[width=0.89\linewidth]{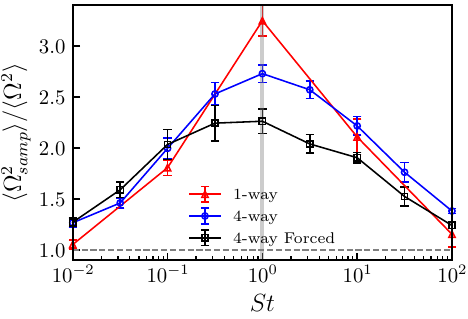}
    \caption{\justifying
    Ratio of enstrophy sampled by particles, $\Omega^2_{\text{samp}}$, to the Eulerian enstrophy, $\Omega^2$, plotted against the Stokes number $St$. Results are presented for 1-way coupling and 4-way coupling with and without external forcing on the particles. We are interested in $St \approx 1$, where light particles preferentially concentrate in vortex filaments (regions of high enstrophy). The 4-way coupling simulations are performed at 1\% volume fraction. For the forced case, the following parameters are used: $a_e = 64 a_{\eta}$ and $\omega = 3.2 \tau_{\eta}^{-1}$.
}
    \label{fig:samp_oo}
\end{figure}

To better understand our \emph{tool}, the Lagrangian forcing, and how it affects Lagrangian statistics, we analyze the preferential concentration of particles with and without external forcing. A way to quantitatively assess preferential concentration is the ratio of enstrophy sampled by particles, $\Omega^2_{\text{samp}}$, to the Eulerian enstrophy, $\Omega^2$. This is shown in \autoref{fig:samp_oo} versus the Stokes number. We focus on $St \sim 1$, where preferential concentration is strongest. The figure shows that both collisions and external forcing reduce this concentration by enabling particles to escape from vortex filaments. While the sampling becomes less strong, it is still far from uniform. This highlights important implications for the Eulerian side: turbulence statistics may be altered through targeted Lagrangian feedback. A key issue is whether preferential sampling of coherent structures plays an essential role in this modification, or if similar effects could be achieved with particles distributed more uniformly.

\begin{figure*}
    \centering
    \includegraphics[width=0.95\linewidth]{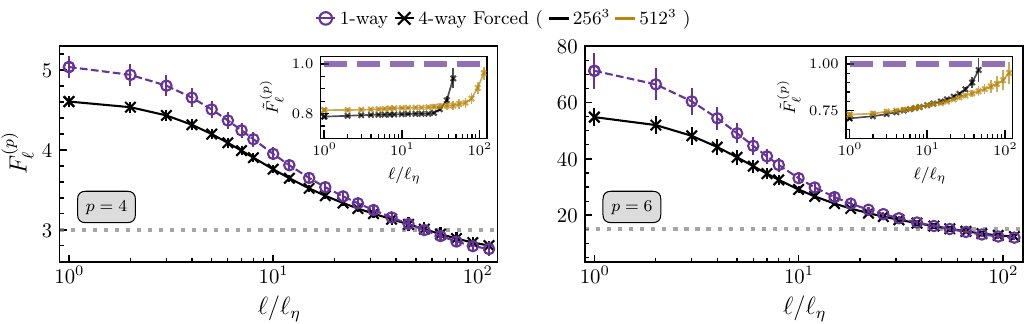}
    \caption{\justifying \textbf{Intermittency suppression.} Eulerian flatness $F_{\ell}^{(p)}$  ($p = 4$ and $p = 6$ considered) for 1-way coupling and 4-way coupling with external forcing acting on the particles. The upper right inset plots show the excess flatness divided by the one with no particles, $\tilde{F}^{(p)} = (F^{(p)} - F^{(p)}_{\text{Gaussian}})/(F^{(p)}_{\text{no particles}} - F^{(p)}_{\text{Gaussian}}))$. The results for the forced cases are presented at two different resolutions: $256^3$ ($Re_{\lambda}=87$) and $512^3$ ($Re_{\lambda}=168$). The 4-way coupling simulations are performed at 1\% volume fraction and with the following external forcing parameters: $a_e = 64 a_{\eta}$ and $\omega = 3.2 \tau_{\eta}^{-1}$.
    } 
    \label{fig:f46}
\end{figure*}

\emph{Intermittency suppression and resonance} - 
This raises our central question: can we selectively control some degrees of freedom of turbulence via a Lagrangian forcing? More specifically, we are interested in whether this control extends to the level of flow intermittency. For this we look at statistics of the fluid when changing the Lagrangian forcing parameters. Namely, we look at the statistics of longitudinal velocity increments $\delta_\ell u=(\boldsymbol{u}(\boldsymbol{x}+\boldsymbol{\ell})-\boldsymbol{u}(\boldsymbol{x})) \cdot \boldsymbol{\ell} / \ell$, in particular we study the scaling of Eulerian structure functions
\begin{equation}
    S^{(p)}_{\ell} \equiv \langle (\delta_\ell u)^p\rangle\,.
\end{equation}
Intermittency is best measured by the departure of the scaling of the structure functions from Kolmogorov's 1941 prediction, $S^{(p)}_{\ell} \sim \ell^{\zeta_p}$, with $\zeta_p = p/3$ in the inertial range~\cite{benzi_inertial_2010}. One way to quantify this is via flatness or kurtosis which can be defined as a dimensionless ratio between structure functions
\begin{equation}
    F^{(p)}_{\ell} = \frac{S^{(p)}_{\ell}}{\left(S^{(2)}_{\ell}\right)^{p/2}}\,.
\end{equation}
As we increase the order $p$ we are essentially giving more weight to the tails of the PDF of $\delta_\ell u$, highlighting more extreme events.
In \autoref{fig:f46}, we present the Eulerian flatness of orders $p = 4$ and $p = 6$, examining how intermittency varies under the Lagrangian forcing. A clear reduction in flatness throughout the full inertial range is observed when the forcing is applied. Beyond a net decrease, the shape of the flatness also changes systematically. At small scales, the slope becomes slightly flatter and the crossover toward the inertial range shifts to larger \(\ell/\ell_\eta\), indicating a scale-dependent modulation rather than a uniform rescaling. The Taylor-scale Reynolds number remains unchanged within uncertainties between the unforced and forced cases, confirming that the forcing does not alter the large-scale turbulence intensity. For the forced case, we also include in the inset the results obtained at higher resolution ($512^3$, $Re_{\lambda}=168$). The suppression of intermittency is very similar to the lower resolution results, but extends over a broader range of scales, thereby consolidating the conclusions drawn from the $256^3$ simulations.

\begin{figure}[b]
    \centering
    \includegraphics[width=0.94\linewidth]{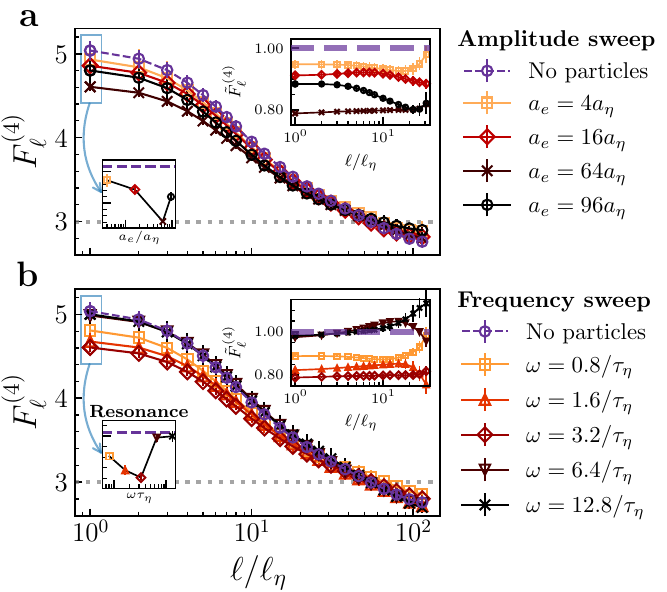}
    \caption{\justifying \textbf{Forcing amplitude and resonance effects.} Eulerian flatness $F_{\ell}^{(4)}$ for (a). Different particle forcing amplitudes $a_e \in [4, 96]a_{\eta}$, at a fixed frequency $\omega = 3.2/\tau_{\eta}$, (b). Different forcing frequencies $\omega \in [0.8, 12.8]/\tau_{\eta}$ at a fixed amplitude $a_e = 64a_{\eta}$. The upper right insets show the excess flatness of selected cases with forced particles divided by the one with no particles, $\tilde{F}^{(4)} = (F^{(4)}-3)/(F^{(4)}_{\text{no particles}}-3)$.}
    \label{fig:sweep}
\end{figure}

For completeness, we have also computed the flatness using the transverse structure functions (see \textit{Supplemental Material}), defined as $S_{\ell}^{\perp(p)} = \langle \left[(\boldsymbol{u}(\boldsymbol{x}+\boldsymbol{\ell}) - \boldsymbol{u}(\boldsymbol{x})) \cdot \hat{\boldsymbol{e}}_{\perp}\right]^p \rangle$, which probe velocity increments perpendicular to the separation vector and are therefore more sensitive to rotational motions associated with high enstrophy regions. The results show that intermittency is also reduced when using the transverse structure functions, although the effect is slightly smaller (about 12–15\%), compared to the longitudinal case. We interpret this as the forcing primarily modifying the strain–vorticity coupling rather than directly damping rotation. We furthermore confirm that the intermittency suppression is largely isotropic, showing only negligible differences between the forcing direction and the other directions (see End Matter).

In \autoref{fig:sweep}, we study how the amplitude and frequency of the Lagrangian forcing influence flatness. Panel~(a) shows a sweep over forcing amplitude, $a_e \in [4, 96]a_{\eta}$, at a fixed frequency $\omega = 3.2/\tau_{\eta}$, while panel~(b) explores a frequency sweep, $\omega \in [0.8, 12.8]/\tau_{\eta}$, at fixed amplitude $a_e = 64a_{\eta}$. 
In the amplitude sweep, increasing the forcing amplitude systematically reduces the flatness, at small and intermediate scales relative to the case without particles, up to a certain optimal value after which the effect decreases. This demonstrates a suppression of intermittency, with the strongest effect observed at $a_e = 64a_{\eta}$. The optimal forcing amplitude can be interpreted through a forcing length $\ell_F = a_e/\omega^2$. For the optimal parameters, $\ell_F \simeq 6\ell_\eta$, which is in line with the order of the vortex filament width $\mathcal{O}(\ell_\eta)$ \cite{Ghira2022,Kang2007,Tanahashi2001,daSilva2011,Ganapathisubramani2008}. 
In the frequency sweep, we find that increasing the frequency enhances intermittency suppression up to $\omega = 3.2/\tau_{\eta}$, suggesting a resonance-like effect where the forcing is most effective when its frequency is tuned to the characteristic timescale of the smallest-scale turbulence, $\tau_{\eta}$. Beyond this frequency, the flatness returns to levels comparable to the unforced case, indicating a loss of phase coherence between the forcing and the flow dynamics. This decorrelation renders the forcing ineffective at higher frequencies. 
The observed resonance was extensively studied in the 1-way coupling regime by Wang et al.~\cite{wang_localization-delocalization_2024}. In that work, this resonance was interpreted as a forced damped-oscillator response of the trapped light particles, leading to a transition between localized and delocalized regimes. Our results confirm the same mechanism in the fully coupled regime, where the resonance together with the momentum exchange produce a measurable modulation of turbulence intermittency.

\begin{figure}[b]
    \centering
    \includegraphics[width=0.88\linewidth]{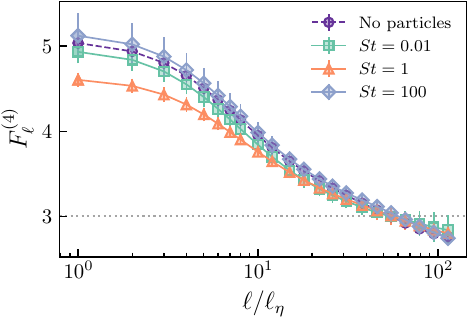}
    \caption{\justifying Fourth-order Eulerian flatness $F^{(4)}_{\ell}$ for different Stokes numbers. The case $St=1$ yields preferential concentration in vortex filaments (see \autoref{fig:samp_oo}), which can be observed to produce significant suppression of intermittency.
    }
    \label{fig:f4_stokes}
\end{figure}

To verify that the observed effects are non-trivial, \autoref{fig:f4_stokes} compares the baseline case without particles to cases with forced particles of varying Stokes numbers: $St = 0.01$ (tracer-like), $St = 1$ (used throughout the paper), and $St = 100$ (heavy-particle-like). The goal is to assess whether the feedback effect depends on preferential concentration in vortex filaments, or if it persists regardless of clustering—implying a trivial response. To isolate the role of particle dynamics, we vary the Stokes number while keeping the density ratio $\beta$ fixed, ensuring equal feedback intensity of the particle-fluid momentum exchange across cases. The results show that for $St = 0.01$ and $St = 100$, where there is virtually no preferential concentration (see \autoref{fig:samp_oo}), the flow remains nearly unchanged from the baseline, indicating that Lagrangian forcing is ineffective without significant clustering. In contrast, $St = 1$ shows clear intermittency suppression, confirming that the effect is non-trivial and relies on preferential concentration in high-vorticity regions. 

\begin{figure}[tb]
    \centering
    \includegraphics[width=0.88\linewidth]{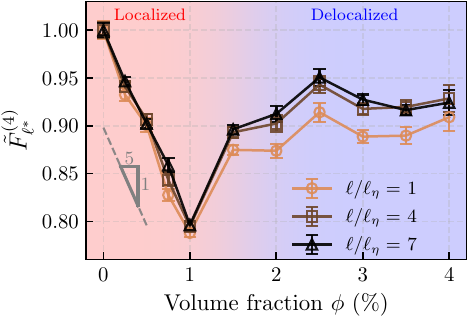}
    \caption{\justifying Fourth-order Eulerian flatness of the case with forced particles divided by the one with no particles, $\tilde{F}^{(4)} = (F^{(4)}-3)/(F^{(4)}_{\text{no particles}}-3)$, at some fixed $\ell^* = \ell/\ell_{\eta}$ as a function of the volume fraction $\phi (\%)$. The following external forcing parameters are used: $a_e = 64 a_{\eta}$ and $\omega = 3.2 \tau_{\eta}^{-1}$. The localized regime corresponds to vortex filaments being saturated with particles and minimal presence of untrapped ones, while the delocalized regime involves a significant fraction of particles outside filaments, where their collective influence becomes less effective.
    }
    \label{fig:f4_vf}
\end{figure}

Lastly, we examine the effect of particle volume fraction on intermittency suppression. In \autoref{fig:f4_vf}, we plot the normalized flatness $\tilde{F}^{(4)} = (F^{(4)} - 3)/ (F^{(4)}_{\text{no particles}}-3)$ at fixed length scales $\ell^*$, for volume fractions $\phi \in [0, 4.0]\%$.
For $\phi \in [0, 1.0]\%$, $\widetilde{F}^{(p)}$ decreases approximately linearly with increasing $\phi$, showing enhanced intermittency suppression as more particles accumulate within vortex filaments. This trend saturates around $\phi \approx 1\%$, where the filaments are effectively filled, in agreement with what was found recently by de Wit et al.~\cite{dewit2025dynamicssmallbubblesturbulence}.
Beyond this point, $\widetilde{F}^{(4)}$ increases to a plateau. Although the trapped particles remain, the excess, now delocalized, particles interact with the surrounding flow and with particles inside filaments. These interactions may weaken the suppression effect, leading to only a partial recovery of flatness effect at higher volume fractions.

\emph{Conclusions and outlook} –  
We have shown that externally forced light particles can modulate turbulence through a non-trivial Lagrangian feedback mechanism. Using high-resolution four-way coupled simulations, we demonstrated that applying an oscillatory force to particles alters their interaction with coherent structures—most notably vortex filaments—resulting in measurable suppression of turbulence intermittency. This effect is strongest when the forcing frequency resonates with the Kolmogorov timescale of turbulence, when the forcing amplitude sets a characteristic length scale comparable to the width of vortex filaments, and when particles are preferentially concentrated due to their inertia. Our findings indicate that turbulence modulation is not a passive outcome of particle presence but depends critically on the ability of particles to cluster in high-vorticity regions and dynamically interact with the flow.

These findings shed new light on the intricate---but still poorly understood---relationship between vortex filaments and turbulence intermittency~\cite{frisch}. For the first time, we demonstrate that turbulence intermittency can be partially suppressed through a physically realizable mechanism, without any artificial modification of the Navier–Stokes equations~\cite{buzzicotti_statistical_2020}. Here, the modulation emerges naturally from the coupling between externally forced light particles and the surrounding coherent structures, establishing an experimentally accessible route to controlling small-scale turbulence statistics. Future work could investigate optimal forcing strategies (possibly using reinforcement learning or other machine learning techniques), effects in transitional flows where the impact may be even more pronounced, or experimental implementations using magnetically or electrically driven particles. More broadly, this approach opens new avenues for targeted manipulation of turbulence using active embedded agents.

\emph{Acknowledgments} - A.F. acknowledges useful discussions with Fabio Bonaccorso and Michele Buzzicotti. This research was supported by European Union’s HORIZON MSCA Doctoral Networks programme under Grant Agreement No. 101072344, project AQTIVATE (Advanced computing, QuanTum algorIthms and data-driVen Approaches for science, Technology and Engineering), the European Research Council (ERC) under the European Union’s Horizon 2020 research and innovation programme Smart-TURB (Grant Agreement No. 882340). This work is supported by the Netherlands Organization for Scientific Research (NWO) through the use of supercomputer facilities (Snellius) under Grant No. 2025.008. This publication is part of the project “Shaping turbulence with smart particles” with Project No. OCENW.GROOT.2019.031 of the research program Open Competitie ENW XL which is (partly) financed by the Dutch Research Council (NWO).

\emph{Data availability} - data will be made available upon reasonable request.

\bibliography{apssamp}

\clearpage

\onecolumngrid
\vspace{5mm}
\begin{center}
    \large \textbf{End Matter}
\end{center}

\begin{table}[htb]
    \centering
    \caption{\justifying The range of simulation parameters used in the paper. From left to right: $N^3$ is the grid resolution, $L$ the box size, $Re_\lambda$ the Taylor-scale Reynolds number, $\ell_\eta$ the Kolmogorov length scale, $dx$ the grid spacing, $\nu$ the kinematic viscosity, $\epsilon$ the energy dissipation rate, $\tau_\eta$ the Kolmogorov timescale, $dt$ the time step, $T$ the total simulation time, $a_e$ the particle forcing amplitude, $\omega$ the particle forcing frequency, $St$ the Stokes number, $\beta$ the density ratio parameter, $\phi$ the volume fraction of particles, and $D$ the particle diameter.} 
    \resizebox{\columnwidth}{!}{
    \begin{tabular}{ccccccccccccccccc}
        \toprule
        $N^3$ & $L$ & $Re_\lambda$ & $\ell_\eta$ & $dx$ & $\nu$ & $\epsilon$ & $\tau_\eta$ & $dt$ & $T$ & $a_e a_{\eta}^{-1}$ & $\omega\tau_{\eta}$ & $St$ & $\beta$ & $\phi$ & $D$\\
        \midrule
        $256^3$ & $2\pi$ & 87  & $2.45 \times 10^{-2}$ & $2.45 \times 10^{-2}$ & $6.02 \times 10^{-4}$ & $6.02 \times 10^{-4}$ & 1 & $5 \times 10^{-3}$ & 264 & $[3.125, 64]$ & $[0.5, 8.0]$ & $\{0.01, 1, 100\}$ & 3.0 & [0, 4]\% & $0.9l_{\eta}$\\
        \bottomrule
    \end{tabular}
    }
\end{table}

\twocolumngrid

\subsection*{Two-way coupling feedback force derivation}
\begin{figure}[htb]
    \centering
    \includegraphics[width=0.4\linewidth]{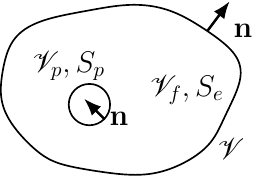}
    \caption{\justifying A particle of volume $\mathscr{V}_p$ and surface $S_p$ is embedded in a fluid volume $\mathscr{V}_f$ with external surface $S_e$. The figure also shows the orientation of the outward normals on $\mathscr{V}_f$.}
    \label{fig:fluid_bubble}
\end{figure}

To derive the expression for the particle feedback force in the two-way coupled system, we consider a single particle embedded in a fluid domain, as illustrated in \autoref{fig:fluid_bubble}. The control volume is split into the fluid region $\mathscr{V}_f$ and the particle region $\mathscr{V}_p$, each bounded by surfaces $S_e$ and $S_p$, respectively.
\paragraph*{Fluid momentum balance.}
The momentum balance for the fluid domain reads:
\begin{equation}
\int_{\mathscr{V}_f} \rho_f \frac{\mathrm{D} \bm{u}}{\mathrm{D} t} \, \mathrm{d} \mathscr{V} = 
\int_{S_e} \bm{\sigma} \cdot \bm{n} \, \mathrm{d} S +
\int_{S_p} \bm{\sigma} \cdot \bm{n} \, \mathrm{d} S \,,
\label{eq:fluid}
\end{equation}
where $\bm{\sigma}$ is the fluid stress tensor and $\bm{n}$ is the unit outward normal vector.
\paragraph*{Particle momentum balance.}
The particle, subject to an external force $\bm{F}_{\text{ext}}$, satisfies:
\begin{equation}
\rho_p \mathscr{V}_p \frac{\mathrm{d} \bm{V}}{\mathrm{d} t} = 
- \int_{S_p} \bm{\sigma} \cdot \bm{n} \, \mathrm{d} S 
+ \bm{F}_{\text{ext}} \,.
\label{eq:particle}
\end{equation}
\paragraph*{Combining the two balances.}
By substituting the common surface term from \autoref{eq:particle} into \eqref{eq:fluid}, we obtain:
\begin{equation}
\int_{\mathscr{V}_f} \rho_f \frac{\mathrm{D} \bm{u}}{\mathrm{D} t} \, \mathrm{d} \mathscr{V} 
- \int_{S_e} \bm{\sigma} \cdot \bm{n} \, \mathrm{d} S  
= - \rho_p \mathscr{V}_p \frac{\mathrm{d} \bm{V}}{\mathrm{d} t} + \bm{F}_{\text{ext}} \,.
\end{equation}
Applying the divergence theorem to convert the surface integral over $S_e$ into a volume integral:
\begin{equation}
\int_{\mathscr{V}_f} \rho_f \frac{\mathrm{D} \bm{u}}{\mathrm{D} t} \, \mathrm{d} \mathscr{V}
- \int_{\mathscr{V}_f} (\nabla \cdot \bm{\sigma}) \, \mathrm{d} \mathscr{V}
= - \rho_p \mathscr{V}_p \frac{\mathrm{d} \bm{V}}{\mathrm{d} t} + \bm{F}_{\text{ext}} \,.
\end{equation}
\paragraph*{Extending to the full domain.}
We now consider the full domain $\mathscr{V} = \mathscr{V}_f + \mathscr{V}_p$ and write:
\begin{equation}
\begin{aligned}
\int_{\mathscr{V}} \rho_f &\frac{\mathrm{D} \bm{u}}{\mathrm{D} t} \, \mathrm{d} \mathscr{V} 
= \int_{\mathscr{V}} \nabla \cdot \bm{\sigma} \, \mathrm{d} \mathscr{V} \\
& + \int_{\mathscr{V}} \left[
\rho_f \frac{\mathrm{D} \bm{u}}{\mathrm{D} t} - 
\rho_p \frac{\mathrm{d} \bm{V}}{\mathrm{d} t} + 
\frac{\bm{F}_{\text{ext}}}{\mathscr{V}_p}
\right] \mathscr{V}_p \delta (\bm{x}-\bm{X}_p(t))
 \, \mathrm{d} \mathscr{V}.
\end{aligned}
\end{equation}
This identifies the particle-induced force density in the Navier–Stokes equations as:
\begin{equation}
\rho_f \mathbf{f}_p(\bm{x}, t) = 
\left[
\rho_f \frac{\mathrm{D} \bm{u}}{\mathrm{D} t} -
\rho_p \frac{\mathrm{d} \bm{V}}{\mathrm{d} t} +
\frac{\bm{F}_{\text{ext}}}{\mathscr{V}_p}
\right] \mathscr{V}_p \delta(\bm{x} - \bm{X}_p(t)) \,.
\end{equation}
To match this to the expression used in the main text, we need to express \autoref{eq:particle} using the Maxey-Riley equations, retaining the Stokes drag, added mass and pressure gradient terms to find
\begin{equation}
\begin{aligned}
\rho_p \mathscr{V}_p \frac{\mathrm{d} \bm{V}}{\mathrm{d} t} =-6\pi r_p \rho_f \nu (\bm{V}-&\bm{u}
) +\tfrac{1}{2} \rho_f \mathscr{V}_p\left(\frac{\mathrm{D} \bm{u}}{\mathrm{D} t}-\frac{\mathrm{d} \bm{V}}{\mathrm{d} t}\right)\\
&+\rho_f \mathscr{V}_p \frac{\mathrm{D} \mathbf{u}}{\mathrm{D} t} + \bm{F}_{\text{ext}},
\label{eq:mr1}
\end{aligned}
\end{equation}
which can be rewritten as
\begin{equation}
\begin{aligned}
\frac{\mathrm{d} \bm{V}}{\mathrm{d} t} = - \overbrace{\frac{\rho_f}{\mathscr{V}_p (\rho_p+\tfrac{1}{2}\rho_f)} 6\pi r_p \rho_f \nu }^{1/\tau_p}&(\bm{V}-\bm{u}) + \overbrace{\frac{3}{2\tfrac{\rho_p}{\rho_f}+1}}^{\beta} \frac{\mathrm{D} \mathbf{u}}{\mathrm{D} t}\\&+ \underbrace{\frac{1}{(\rho_p+\tfrac{1}{2}\rho_f)\mathscr{V}_p} \bm{F}_{\text{ext}}}_{\bm{f}_{E}}.
\end{aligned}
\end{equation}
This identifies $\bm{F}_{\text{ext}}=(\rho_p+\tfrac{1}{2}\rho_f)\mathscr{V}_p \bm{f}_{E}$, yielding the expression used in the main text
\begin{equation}
\mathbf{f}_p(\bm{x}, t)=\biggr[\frac{\mathrm{D} \bm{u}}{\mathrm{D} t}  - \frac{\rho_p}{\rho_f}\frac{\mathrm{d} \bm{V}}{\mathrm{~d} t} + \underbrace{\left(\frac{\rho_p}{\rho_f}+\frac{1}{2}\right)}_{3/(2\beta)}\bm{f}_{E}\biggr] \mathscr{V}_p \delta(\bm{x}-\bm{X}_p(t)) .
\end{equation}

\subsection*{Decaying turbulence}
\begin{figure}[H]
    \centering
    \includegraphics[width=0.89\linewidth]{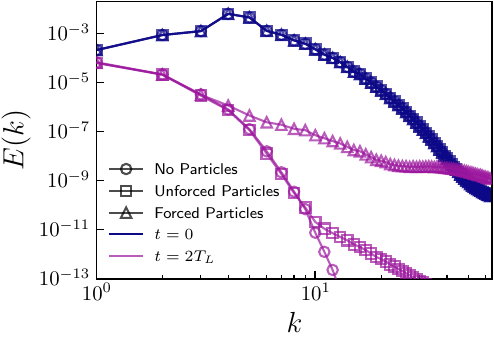}
    \caption{\justifying Energy spectra at $t = 0$ and $t = 2T_L$ for decaying turbulence without particles, with unforced particles and with forced particles. $T_L$ is the integral time scale.}
    \label{fig:energy_decaying}
\end{figure}

A substantial portion of the literature on point-particles embedded in a fluid with two-way coupling has focused on decaying turbulence~\cite{elghobashi_direct_1992, elghobashi_two-way_1993, druzhinin_decay_1999, ferrante_physical_2003, bosse_small_2006, horwitz_two-way_2020}. Studying two-way coupling in the decaying regime offers several practical and conceptual advantages. First, it allows the use of the conventional $2/3$-dealiasing rule without significantly attenuating the particle-to-fluid feedback. Second, the lack of energy injection makes the relative influence of the particle feedback more pronounced. Finally, the decaying setup avoids the assumption that the external Eulerian forcing does not contribute to the pressure gradient force in the derivation of particle phase equations and the particle back-reaction on the fluid. For completeness, and to maintain continuity with prior work, we present here results for decaying (i.e., unforced) turbulence including four-way coupling and external forcing on the particles. In \autoref{fig:energy_decaying}, we show the energy spectra at the initial time $t = 0$ and after two integral time scales have passed, $t = 2 T_L$, for the baseline case without particles, and for unforced and forced particles. It is clearly seen that both the particle back-reaction alone and its combination with external forcing act as effective sources of energy, significantly reshaping the spectrum compared to the baseline case.

\subsection*{Directional response to the particle forcing}
\begin{figure}[H]
    \centering
    \includegraphics[width=0.97\linewidth]{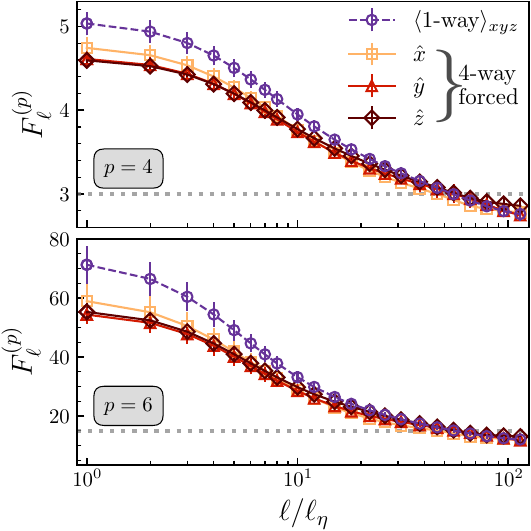}
    \caption{\justifying
    Longitudinal flatness $F_\ell^{(p)}$ for $p=4$ (top) and $p=6$ (bottom)
    with four–way coupling and external particle forcing, shown separately for
    increments along $\hat{\bm e}_x$, $\hat{\bm e}_y$,
    and $\hat{\bm e}_z$. The dashed curve is the 1–way reference
    averaged over directions.}
    \label{fig:f46_dirs}
\end{figure}

The external particle forcing enters with a preferred direction, $\bm{f}_E=a_e\sin(\omega t)\,\hat{\bm e}_x$, so we ask whether it imprints an anisotropic signature on the Eulerian velocity increments. We compute directional longitudinal structure functions using increments along each Cartesian axis, $\delta_\ell u_\alpha=[\bm u(\bm x+\ell\hat{\bm e}_\alpha)-\bm u(\bm x)]\!\cdot\!\hat{\bm e}_\alpha$ with $\alpha\in\{x,y,z\}$, and evaluate the flatness $F_\ell^{(p)}=S^{(p)}_\ell/(S^{(2)}_\ell)^{p/2}$ for $p=4,6$. \autoref{fig:f46_dirs} shows the forced, four–way–coupled case componentwise, compared with the 1–way reference (xyz–averaged).

A weak but systematic ordering is present: the forcing–aligned component ($\hat{\bm e}_x$) shows the smallest departure from the reference, while $\hat{\bm e}_y$ and $\hat{\bm e}_z$ are slightly more affected. This modest anisotropy appears only in higher–order moments (as the ones analyzed here), indicating that it is a subtle, small–scale effect.

Should we expect any directional difference at all? In principle, the forcing selects the $x$–axis and therefore breaks isotropy, yet several processes act to smear out this preference before it reaches the
Eulerian statistics. Particles exchange momentum with the fluid along $\hat{\bm e}_x$ in a time–periodic fashion, but the divergence–free projection of the feedback redistributes this injected momentum among all components, spreading part of it into transverse motions. Moreover, since the light particles interact with the flow mainly through vortex filaments of random orientation, the local momentum exchange is continually reoriented, so the feedback should become nearly isotropic once averaged over many events. The small residual anisotropy observed in \autoref{fig:f46_dirs} possibly stems from particles that are not fully trapped in filaments and thus retain some memory of the forcing direction. Overall, the effect remains very weak and does not qualitatively alter the isotropy of the flow. This once more underpins that the suppression of intermittency is a non-trivial effect, related to the role of vortex filaments in turbulence intermittency, rather than being an artifact of the small-scale forcing.

\pagebreak

\end{document}


\title{Supplemental Material:\\Intermittency suppression in turbulence via forced light particles}

\maketitle

\section{Transverse structure functions}

\begin{figure}[h]
    \centering
    \includegraphics[width=1.0\linewidth]{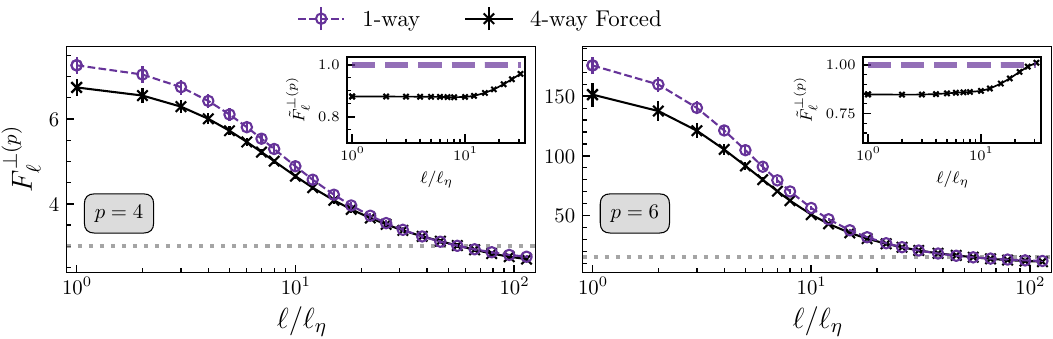}
    \caption{Eulerian transverse flatness $F_{\ell}^{\perp(p)}$  ($p = 4$ and $p = 6$ considered) for 1-way coupling and 4-way coupling with external forcing acting on the particles. The upper right inset plots show the excess flatness divided by the one with no particles, $\tilde{F}^{\perp(p)} = (F^{\perp(p)} - F^{(p)}_{\text{Gaussian}})/(F^{\perp(p)}_{\text{no particles}} - F^{(p)}_{\text{Gaussian}}))$. The gray dashed lines correspond to the Gaussian values. The 4-way coupling simulations are performed at 1\% volume fraction and with the following external forcing parameters: $a_e = 64 a_{\eta}$ and $\omega = 3.2 \tau_{\eta}^{-1}$.
    }     
    \label{fig:fperp}
\end{figure}

To complement the analysis of the longitudinal velocity increments presented in the main text, we also examined the transverse structure functions, which probe the velocity increments perpendicular to the separation vector:
\begin{equation}
S_{\ell}^{\perp(p)} = \big\langle [(\bm{u}(\bm{x}+\ell\hat{\bm{e}}_{\perp})-\bm{u}(\bm{x}))\!\cdot\!\hat{\bm{e}}_{\perp}]^{p} \big\rangle\,.
\end{equation}
These are more sensitive to rotational motions and therefore to regions of high enstrophy, where the light particles preferentially concentrate.
\autoref{fig:fperp} shows the corresponding Eulerian flatness $F_{\ell}^{\perp(p)} = S_{\ell}^{\perp(p)}/(S_{\ell}^{\perp(2)})^{p/2}$ for $p=4$ and $p=6$, comparing the reference one-way coupled case with the four-way coupled one where there is additionally the external oscillatory forcing acting on the particles. Partial intermittency suppression is also observed in the transverse increment, although the effect is weaker than in the longitudinal case: the non-Gaussian contribution decreases by approximately 12\% for $p=4$ and 15\% for $p=6$, compared with 20-25\% in the longitudinal increments. Because the transverse statistics emphasize rotational fluctuations, this behavior suggests that the particle forcing does not directly weaken vortex filaments' rotation, but rather modulated how the vortex filaments exchange energy with the surrounding strain field. In this sense, the intermittency reduction originates mainly from a modified strain-vorticity coupling rather than from a direct damping of rotation intensity.

We can use these results to gain some insight into the elusive link between vortex filaments and intermittency. The comparatively weak change in the transverse statistics hints that vortex filaments may serve less as the origin of intermittency than its scaffolding, coherent structures through which the surrounding strain field concentrates and releases fluctuations. The forcing appears to modulate this exchange rather than the filaments themselves, suggesting that intermittency emerges from this dynamic tension between rotation and strain, not from either element in isolation.